\documentclass[showpacs,aps,prl,twocolumn,groupedaddress,amsmath,amssymb]{revtex4}

\usepackage{graphicx}
\usepackage{dcolumn}
\usepackage{bm}
\usepackage{subfig}

\begin{document}

\title{The Emergence of El-Ni\~{n}o as an Autonomous Component in the Climate Network}

\author{A. Gozolchiani}
\affiliation{
   Minerva Center and Department of Physics, Bar Ilan University, Ramat Gan, Israel.
}%
\email{avigoz@google.com}

\author{K. Yamasaki}%
  \affiliation{Tokyo University of Information Sciences, Chiba, Japan}

\author{S. Havlin}
\affiliation{
   Minerva Center and Department of Physics, Bar Ilan University, Ramat Gan, Israel.
}%
\date{\today}
\begin{abstract}
We construct and analyze a climate network which represents the interdependent structure of the climate in different geographical zones and find that the network responds in a unique way to El-Ni\~{n}o events. Analyzing the dynamics of the climate network shows that when El-Ni\~{n}o events begin, the El-Ni\~{n}o basin partially loses its influence on its surroundings. After typically three months, this influence is restored while the basin loses almost all dependence on its surroundings and becomes \textit{autonomous}. The formation of an autonomous basin is the missing link to understand the seemingly contradicting phenomena of the afore--noticed weakening of the interdependencies in the climate network during El-Ni\~{n}o and the known impact of the anomalies inside the El-Ni\~{n}o basin on the global climate system. 
\end{abstract}


\maketitle

It was recently suggested that climate fields such as temperature and geopotential height at a certain pressure level can be represented as a climate network~\cite{whatdonetwor}. In this network, different regions of the world are represented as nodes which communicate by exchanging heat, material, and by direct forces. These interactions are represented by the links of the climate network. Interactions between two nodes may also exist due to processes which take place outside the atmospheric pressure level or through interactions with the ocean and the lands. Each link is quantified by a weight based on measures of similarity between the time series (e.g. correlations) of the corresponding individual nodes (see \cite{exper_tele} for an experimental evidence for the relations between heat exchange and synchronized fluctuations of the temperature field).

Recent studies~\cite{survival,tsoniselnino,EPLgozyamga} show that many links in the climate network break during El-Ni\~{n}o events. The climate network contains several types of links, that have different levels of responsiveness to El-Ni\~{n}o. From the maps of Tsonis and Swanson~\cite{tsoniselnino}, one can locate the responsive nodes (the nodes attached to the most responsive links) to be in the pacific El-Ni\~{n}o Basin (ENB)~\cite{bjerknes69}. These maps together with the tremendous impact of the El-Ni\~{n}o Southern Oscillations on world climate, suggest that ENB have a unique dynamical role in the dynamics of the climate network. Indeed ENB has unique topological properties. Its connectivity and clustering coefficient fields studied by Donges et. al.~\cite{donges_mi} can be distinguished from their surrounding by their particularly higher values. Also, the betweenness centrality field~\cite{donges} in ENB is very low. However, the dynamics related to the interaction of ENB with its surroundings and the origin of its unique features are still not known~\cite{storchbook722}.  

In this Letter we follow the dynamics of the climate network in time, between the years 1979 and 2009 where eight El-Ni\~{n}o events took place. We identify a cluster around ENB that shows a clear \textit{autonomous} behavior during El-Ni\~{n}o epochs, and determine the dynamics of its interactions with the surroundings. We also find an epoch of a decreased influence of ENB on the surroundings, typically three months before the emergence of the \textit{autonomous} behavior. Our findings resolve the seemingly contradictory situation of decreased interactions of ENB with its surroundings on one hand, as explored in previous works~\cite{survival,tsoniselnino,EPLgozyamga}, and the known influence of ENB on world climate on the other hand. We find that only the influence of the surroundings on the ENB region is significantly reduced, creating, apparently, a dynamical autonomous source within this region which influences its surroundings. 

The adjacency matrix in our climate network analysis is based on a similarity measure between time series of temperatures (after removal of the annual trend) at the pressure level of 1000mb covering the last 30 years. In the current work we pick 726 nodes from the Reanalysis II grid~\cite{noaa}, such that the globe is covered approximately homogeneously. The similarity measure $W_{l,r}^{y}$ we use is defined as follows. First we identify the value of the highest peak of the absolute value of the cross covariance function. Then we subtract the mean of the cross covariance function and divide the difference by its standard deviation (see ~\cite{survival,EPLgozyamga} for further details). The indices $l$ and $r$ represent two nodes, and $y$ is the beginning date of a snapshot of the network, measured over 365 days with extended period of 200 days of shifts needed for evaluating the cross covariance. The time shift of the highest peak of the cross covariance function from zero time shift is denoted $\theta^{y}_{l,r}$. The sign of  $\theta^{y}_{l,r}$ stands for the dynamical ordering of $l$ and $r$ (and hence $\theta^{y}_{l,r}=-\theta^{y}_{r,l}$). In directed graph theory terms, when $\theta^{y}_{l,r}>0$ the link is regarded as outgoing from node $l$ and incoming to node $r$. Until now, such ordering or direction were not considered for constructing climate networks. Our method, thus, enables to distinguish between \textit{in--} and \textit{out--} links.

The adjacency matrix of a weighted directed climate network is defined :

\begin{eqnarray}
A^{y}_{l,r}=\left(1-\delta_{l,r}\right)\Theta\left(\theta^{y}_{l,r}\right)W^{y}_{l,r}
\label{eq:A_def}
\end{eqnarray}

whereas $\Theta(x)$ is the Heaviside function.    

The \textit{in--} and \textit{out--} weighted degrees are defined :

\begin{subequations}
\label{eq:IO_def}
\begin{equation}
I^{y}_{l}=\sum_{r}A^{y}_{r,l}
\end{equation}
\begin{equation}
O^{y}_{l}=\sum_{r}A^{y}_{l,r}
\end{equation}
\end{subequations}

The $I$ and $O$ fields represent the level of the dependence of a node on its surrounding, and the level of its influence on the surrounding, respectively. 

The total weighted degree (also denoted ``vertex strength''~\cite{airports}) of a node $l$ is defined as :
\begin{eqnarray}
D^{y}_{l}=\sum_{r}\left(1-\delta_{\theta_{l,r},0}\right)(A^{y}_{l,r}+A^{y}_{r,l})+\sum_{r}\delta_{\theta_{l,r},0}A^{y}_{l,r}
\label{eq:D_def}
\end{eqnarray}

\begin{figure}
\centering
\captionsetup[subfloat]{position=top}
\subfloat[$D$--field]{\label{fig:tot_deg}\includegraphics[width=0.25\textwidth,angle=270]{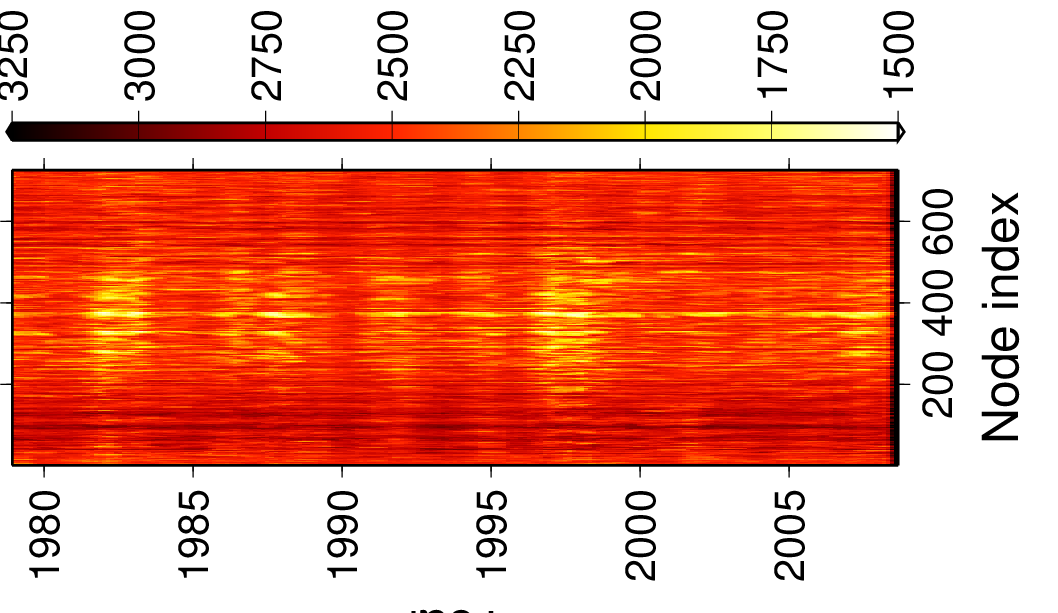}}
\subfloat[$I$--field]{\label{fig:in_clust}\includegraphics[width=0.25\textwidth,angle=270]{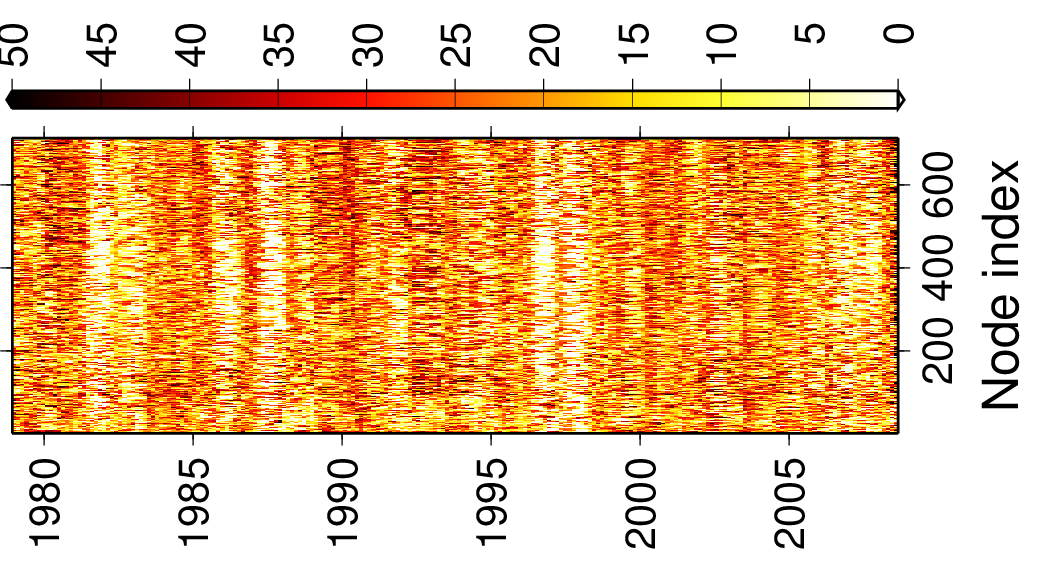}}
\subfloat[$O$--field]{\label{fig:out_clust}\includegraphics[width=0.25\textwidth,angle=270]{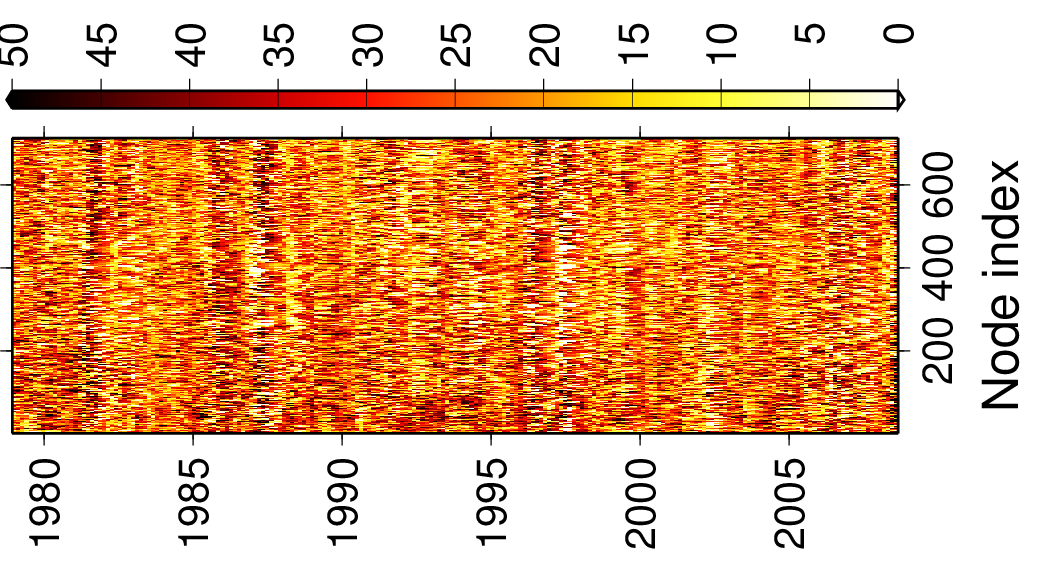}}
\caption{\label{fig:outintot} Weighted degrees as a function of time (y axis) and the node index (x axis). (a) Total weighted degree $D^{y}_{l}$. (b)  The microscopic contributions to the weighted incoming degree of ${\bf C}$, $I^{y}_{\bf C}$. (c) The microscopic contributions to the weighted outgoing degrees, $O^{y}_{\bf C}$. One should bare in mind that each point is compiled from records of 565 days : 365 days + 200 days of shifts. The representative point in all figures for each 565 period is the beginning date of the period}
\end{figure}

Fig.~\ref{fig:tot_deg} shows the time dependence of $D$ for each node in the network. Two main observations can be clearly seen from this figure. First, weighted degrees of the nodes yield an extremely persistent quantity~\cite{fn1}, and different geographical regions have typical values over time. A second notable pattern is the horizontal bright stripes that appear during El-Ni\~{n}o events. The second feature further supports the finding~\cite{survival,tsoniselnino,EPLgozyamga} that many links of the climate network break during El-Ni\~{n}o events. 

Inside the mid range of node indices, between 300 and 500, there is a group of nodes, ${\bf C}$, that have lower values at any time, and are specifically distinct during El-Ni\~{n}o (Fig.~\ref{fig:tot_deg}). These low values and their distinct response to El-Ni\~{n}o, as we will explicitly show, are mainly due to \textit{in--} links. When studying the \textit{in--} and \textit{out--} links separately, we find that the spatial distribution of the \textit{in--} weighted degrees $I_{l}^{y}$ is much broader compared to the distribution of the \textit{out--} weighted degrees $O_{l}^{y}$, at any time $y$, typically by 10 to 20 percent. The 90-th percentile (statistics is collected over time) yields a deviation of 30 percents. As shown below, this difference between $I$ and $O$ is mainly contributed  by the group ${\bf C}$ discussed above. In order to follow the different dynamical behavior of $I$ and $O$, we apply the following approach.

\begin{figure}
\centering
\subfloat[$\left<I_{l}^{y}\right>_{y\in El-Nino}$]{\label{fig:indisten}\includegraphics[width=0.15\textwidth,angle=270]{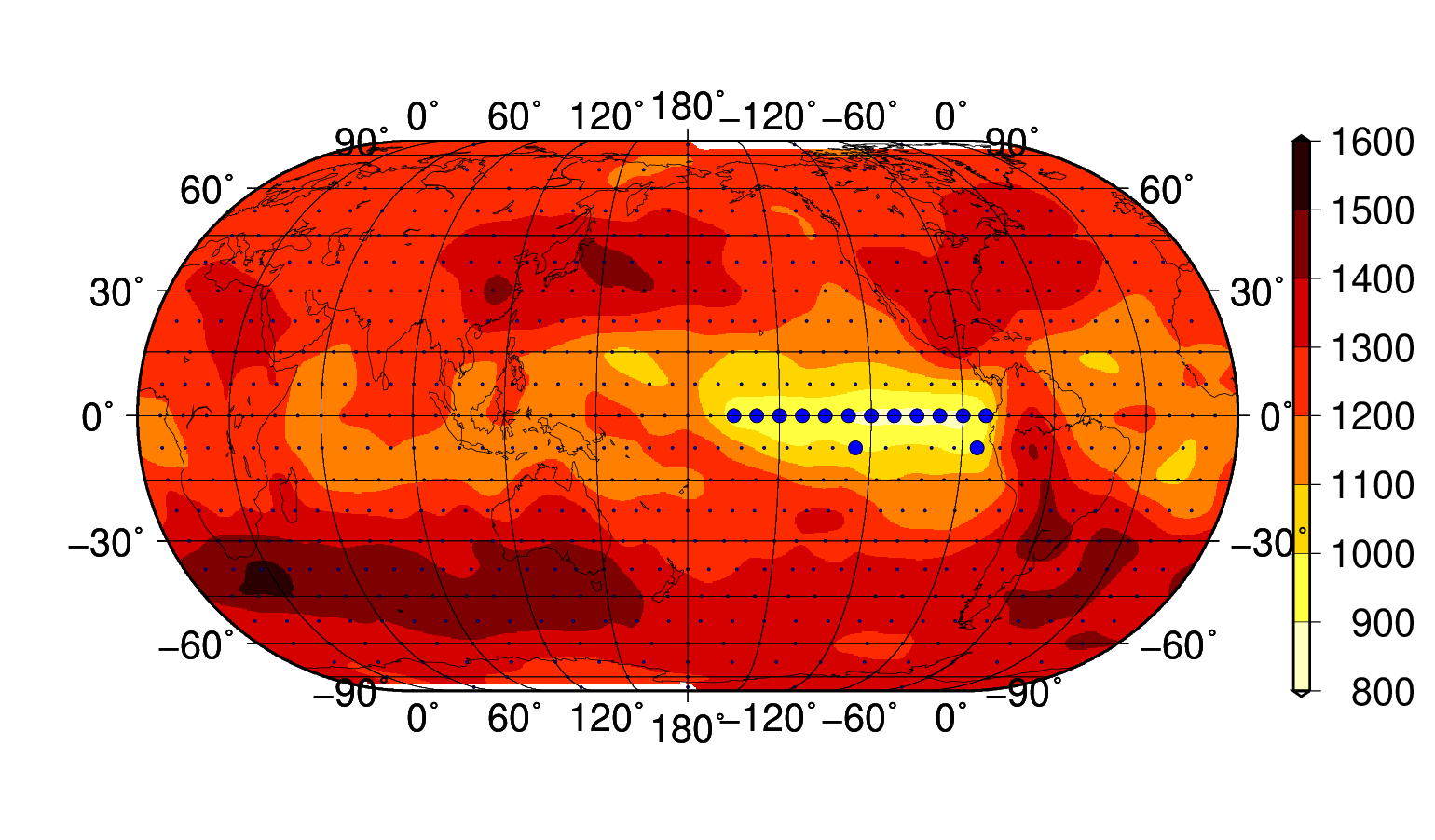}}
\subfloat[$\left<O_{l}^{y}\right>_{y\in El-Nino}$]{\label{fig:outdisten}\includegraphics[width=0.15\textwidth,angle=270]{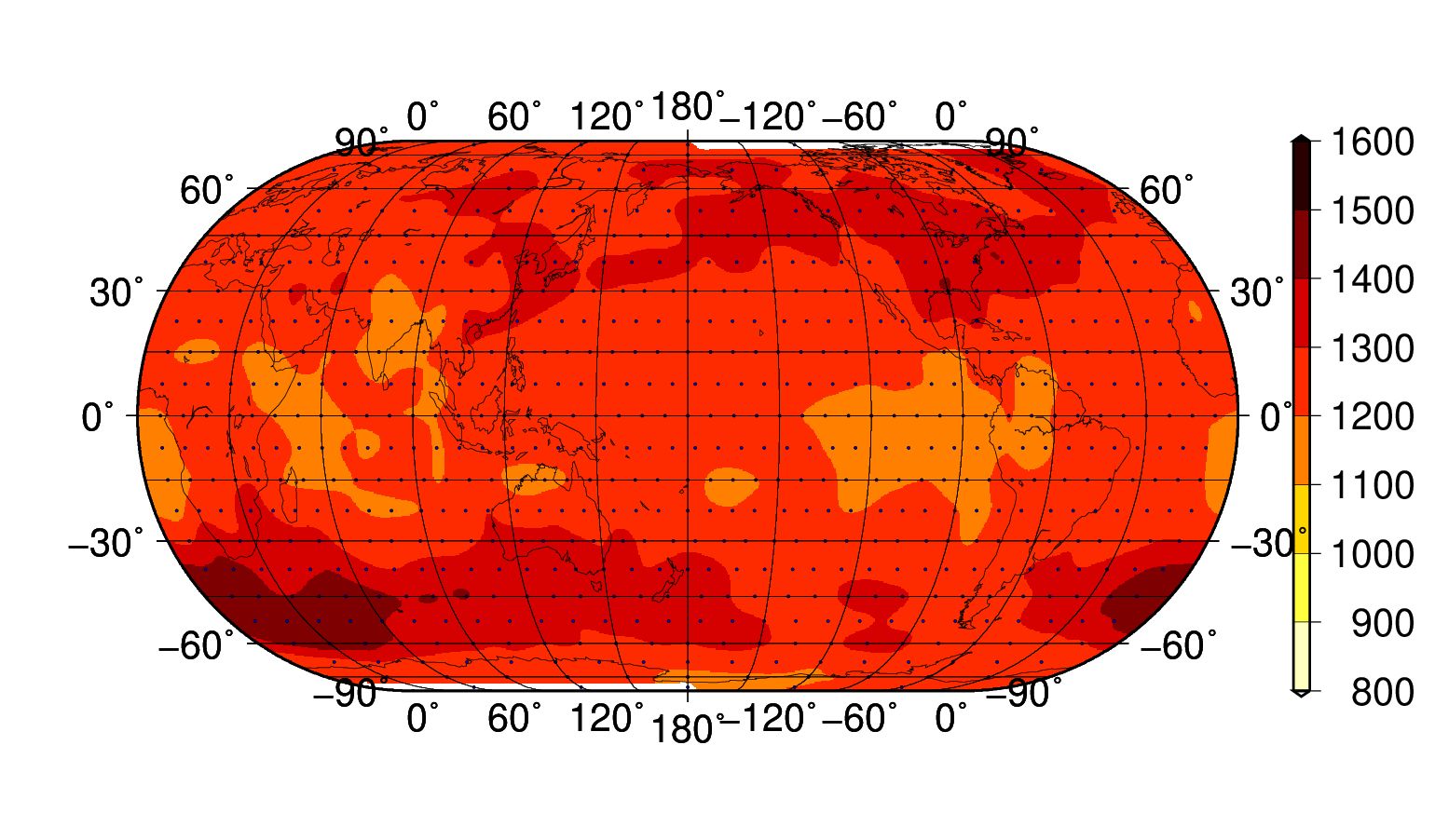}}
\caption{\label{fig:inoutdisten}Spatial distribution of the weighted degrees of nodes (\textit{in} and \textit{out}) averaged over El-Ni\~{n}o times. The nodes of the network are shown by small dots. The nodes belonging to  ${\bf C}$ are shown on the weighted \textit{in--}degree map as larger blue circles}
\end{figure}

We consider only links that are related directly to the El-Ni\~{n}o dynamics by focusing on the group of nodes ${\bf C}$ that have extremely low weighted degrees during El-Ni\~{n}o events:   

\begin{eqnarray}
  {\bf C}\equiv\left\{l\ |\ \left<D^{y}_{l}\right>_{y\in EN}\leq T\right\}
  \label{eq:in_thresh}
\end{eqnarray} 

where $T=2150$ is a threshold~\cite{fn2}, and the triangular brackets stand for averaging over all El-Ni\~{n}o events. The group ${\bf C}$ which includes 14 nodes is part of ENB, and is notably located in the eastern equatorial part of ENB, which is known to have a large scale upwelling of cold ocean water, yielding a ``cold tongue''~\cite{bjerknes69} which deforms during El-Ni\~{n}o (see Fig.~\ref{fig:indisten} for its location).

\begin{figure}
\includegraphics[width=0.41\textwidth]{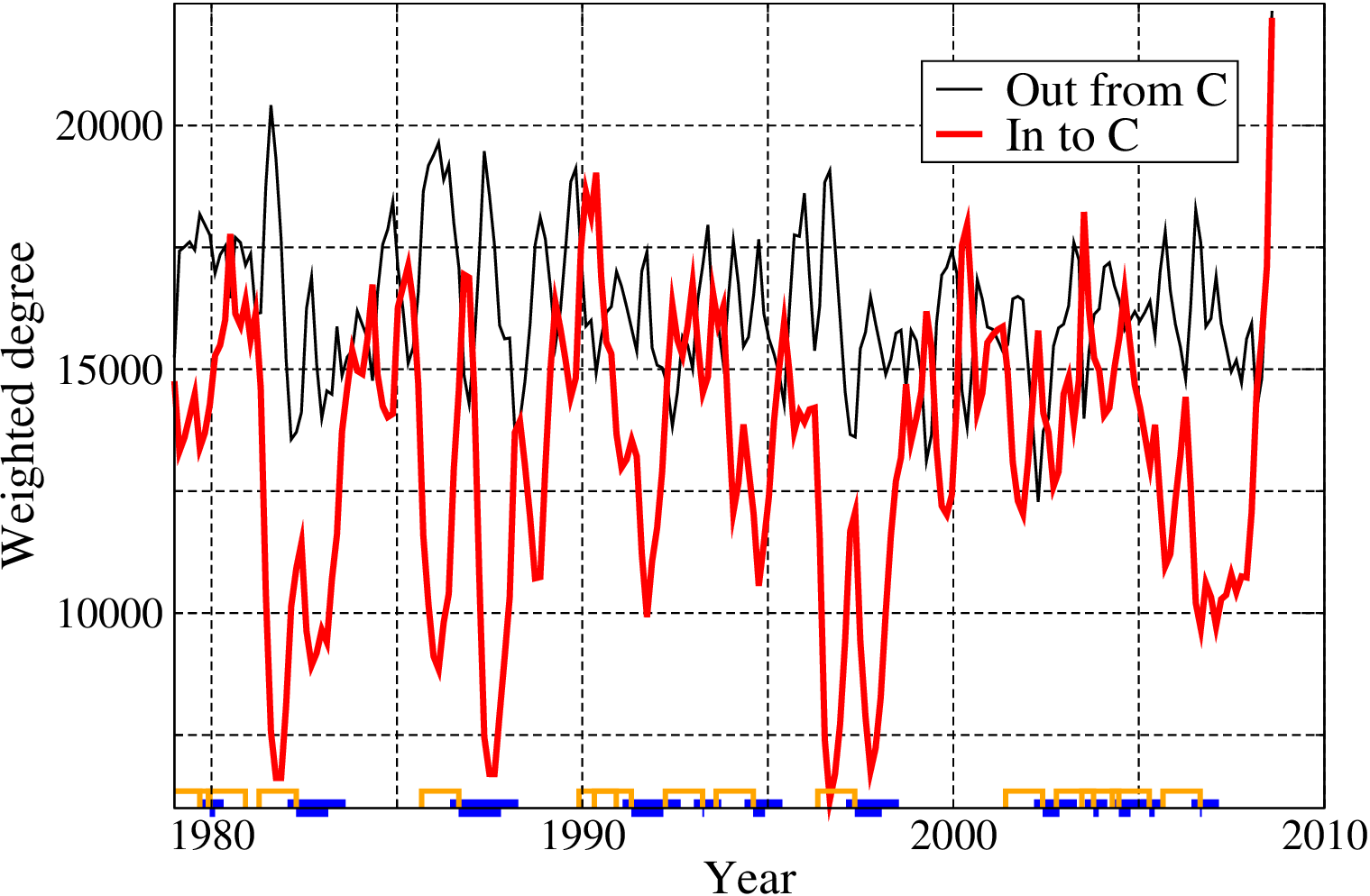}
\caption{\label{fig:sum_in_out_clust} Plot of the weighted \textit{in--} and \textit{out--} degrees of ${\bf C}$, $I^{y}_{\bf C}$ and $O^{y}_{\bf C}$,  respectively. The blue marker along the x--axis shows the periods in which the NINO3.4 (see e.g.~\cite{nino34}) mean temperature exceeds $0.4^{o}C$, which is one of the criterions indicating an El-Ni\~{n}o event. The orange markers show the extended periods in which our analysis is influenced by the events because of the 365 days integration of the cross covariance function.}
\end{figure}

We find that measuring the dynamics of the interactions of the ${\bf C}$ region with its surroundings yields a sensitive tool to quantify the responses of $I$ and $O$ to El-Ni\~{n}o events. One can define $I^{y}_{\bf C}$ and $O^{y}_{\bf C}$, the \textit{in--} and \textit{out--} weighted degree of ${\bf C}$, respectively, in the same manner we defined the weighted degrees for nodes. We consider only links between pairs of nodes where one node belongs to ${\bf C}$ and the other does not. Fig.~\ref{fig:sum_in_out_clust} shows the time dependence of these two variables. As seen in the figure, the responses of the \textit{in--} and \textit{out--} degrees to El-Ni\~{n}o seem to be anti-correlated. When the \textit{in--}degree drops, the \textit{out--}degree slightly increases. The interpretation is that the nodes inside ${\bf C}$ lose large part of their dependence on the surrounding, while slightly increase their influence outside, thus becoming significantly more \textit{autonomous} during El-Ni\~{n}o. Note that the excess of \textit{out--} links over \textit{in--} links is evident also in normal times. The detailed space-dependent picture, however, is even richer, as discussed below. 

The spatial dependence of $I^{y}_{\bf C}$ and $O^{y}_{\bf C}$ is explored by what we call the ``microscopic'' contributions from each one of the 712 nodes that \textit{do not} belong to ${\bf C}$. These microscopic contributions can be visualized (Figs.~\ref{fig:outintot}\subref*{fig:in_clust},\subref*{fig:out_clust}) as they evolve in time. 

El-Ni\~{n}o events influence $I$ and $O$ in a very different way, having a rather peculiar phase locking. Strong El-Ni\~{n}o events weaken the links, yielding smaller microscopic contributions to both  $I^{y}_{\bf C}$ and $O^{y}_{\bf C}$ (white horizontal stripes in Figs.~\ref{fig:outintot}\subref*{fig:in_clust},\subref*{fig:out_clust}), just like Fig.~\ref{fig:tot_deg}. However, the \textit{in--}links are significantly more vulnerable to El-Ni\~{n}o compared to the \textit{out--}links. Moreover, before each major event of weakening $I$-field, there is a short, weaker response of weakening of the $O$-field. After the weakening epoch of both fields the $O$ field recovers, and thereafter the $I$ field  recovers as well, leading to a full oscillation of both fields~\cite{fn3}. It is possible to summarize the situation by saying that the ${\bf C}$ nodes lose their autonomous power at the beginning of an El-Ni\~{n}o event, but thereafter, during the event, it recovers and becomes significantly more autonomous than the average. Despite the prominent difference during El-Ni\~{n}o between their dynamics, the $I$ and $O$ fields exhibit a remarkable static  mirror similarity. Both fields are found to be highly asymmetrical with respect to the equator, at all times. We find that the \textit{in--} links (going \textit{into} ${\bf C}$) mainly originate from the northern latitudes, while the \textit{out--} links (\textit{out from} ${\bf C}$) mainly target towards the southern latitudes. 

\begin{figure}
\includegraphics[width=0.40\textwidth]{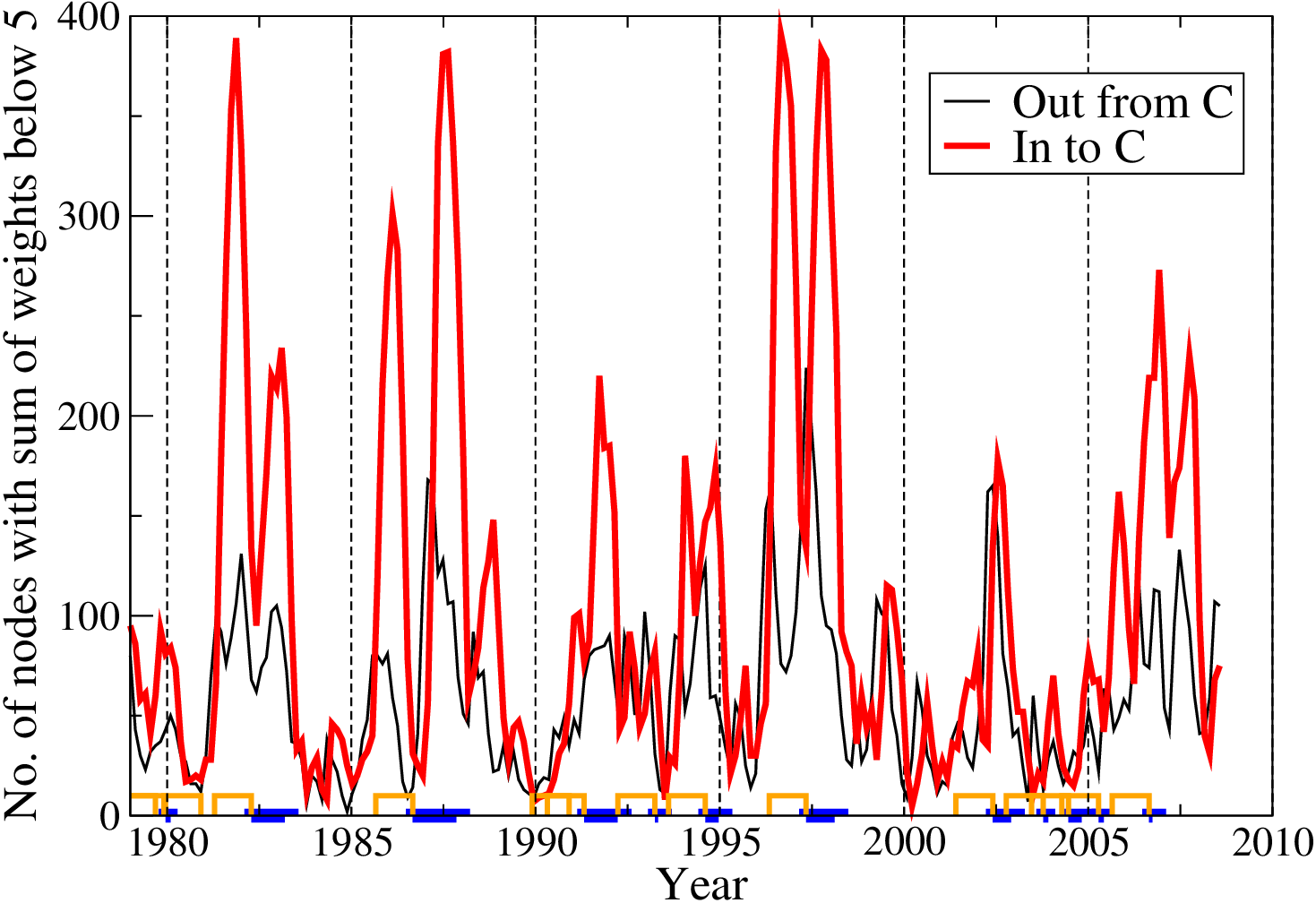}
\caption{\label{fig:in_out_clust} Number of nodes outside ${\bf C}$ having sum of weights below 5 directing towards ${\bf C}$ (in), $I^{*}$, or from ${\bf C}$ (out),$O^{*}$. The blue and orange markers are described in the previous figure}
\end{figure}

\begin{figure}
\includegraphics[width=0.40\textwidth,clip,trim=0 0 0 2]{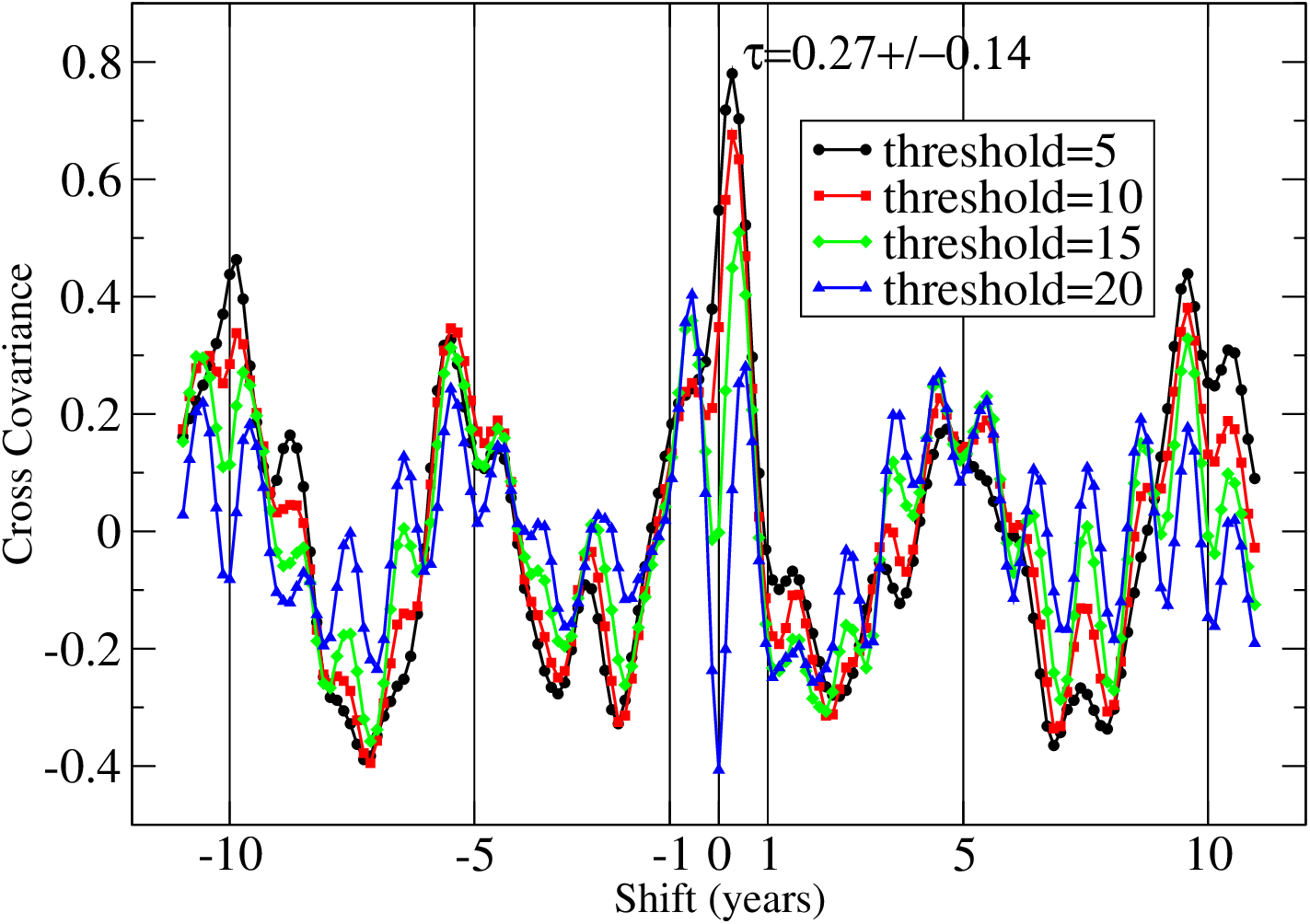}
\caption{\label{fig:x_in_out_clust} Cross covariance of the number of links below various thresholds going into and out from cluster ${\bf C}$ (\textit{in} and \textit{out} respectively, Fig.~\ref{fig:in_out_clust}). In the positive direction of the x-axis, the number of \textit{in--} links lags in time after the \textit{out--} links. }
\end{figure}

In order to detect the weak response of $O^{y}_{\bf C}$, we next consider specific ``microscopic'' contributions that have the lowest weight values, disregarding the rest of the field. This filters out noise that is not related to the observed slight weakening. We count (Fig.~\ref{fig:in_out_clust}) the number of nodes having microscopic contribution to the weighted \textit{in--} and \textit{out--} degrees under a weight threshold of 5, defined as $I^{*}$ and $O^{*}$ respectively. We will show (Fig.~\ref{fig:x_in_out_clust}) that there is little sensitivity to this threshold value. Thus, a rise of $O^{*}$ (red) in Fig.~\ref{fig:in_out_clust} corresponds to nodes losing their dependence on ${\bf C}$. Shortly after such events we see a significant rise of the $I^{*}$ (black) that corresponds to nodes losing their influence on ${\bf C}$. As is clearly seen in Fig.~\ref{fig:in_out_clust}, the $I^{*}$ and $O^{*}$ oscillations begin at the onset of El-Ni\~{n}o events. 

Next, we extract the time lag between the two coherent oscillations of $I^{*}$ and $O^{*}$ using the cross covariance function. In Fig.~\ref{fig:x_in_out_clust} we show the cross covariance between the two curves of Fig.~\ref{fig:in_out_clust}, for several threshold values. The peak of the cross covariance function shows that $O$ lags after $I$ in $100\pm 50$ days~\cite{fn5}. The sharp correlation between the number of ``weak'' nodes in both directions (\textit{in} and \textit{out}) as well as the time lag is persistent up to a threshold value of 15. The other peaks of the cross correlation function are related to the quasi-periodicity of the El-Ni\~{n}o southern oscillations. 

\begin{figure}
\centering
\subfloat[$\left<I_{l}^{y}\right>_{y\not\in El-Nino}$]{\label{fig:indistnonen}\includegraphics[width=0.15\textwidth,angle=270]{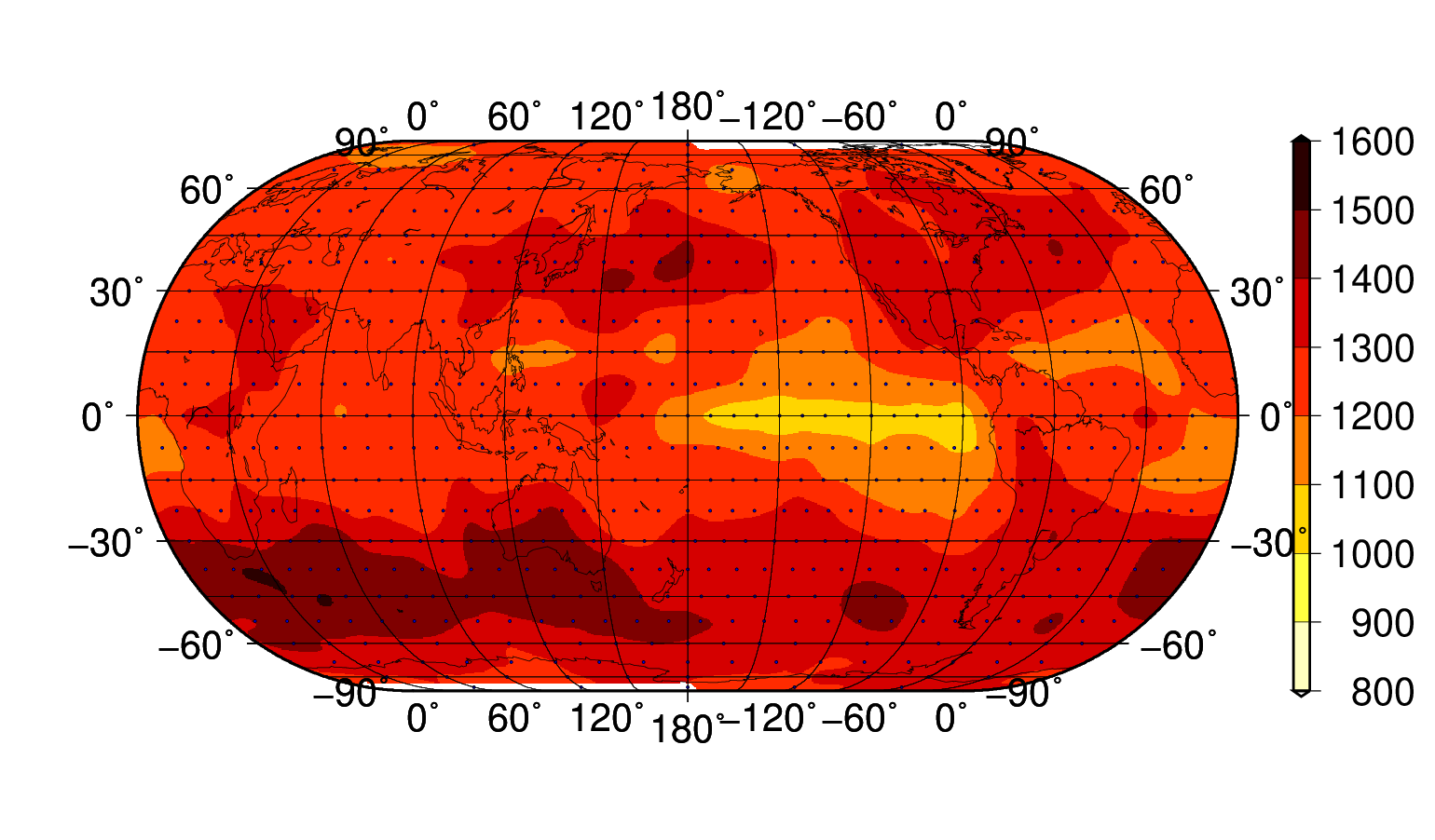}}
\subfloat[$\left<O_{l}^{y}\right>_{y\not\in El-Nino}$]{\label{fig:outdistnonen}\includegraphics[width=0.15\textwidth,angle=270]{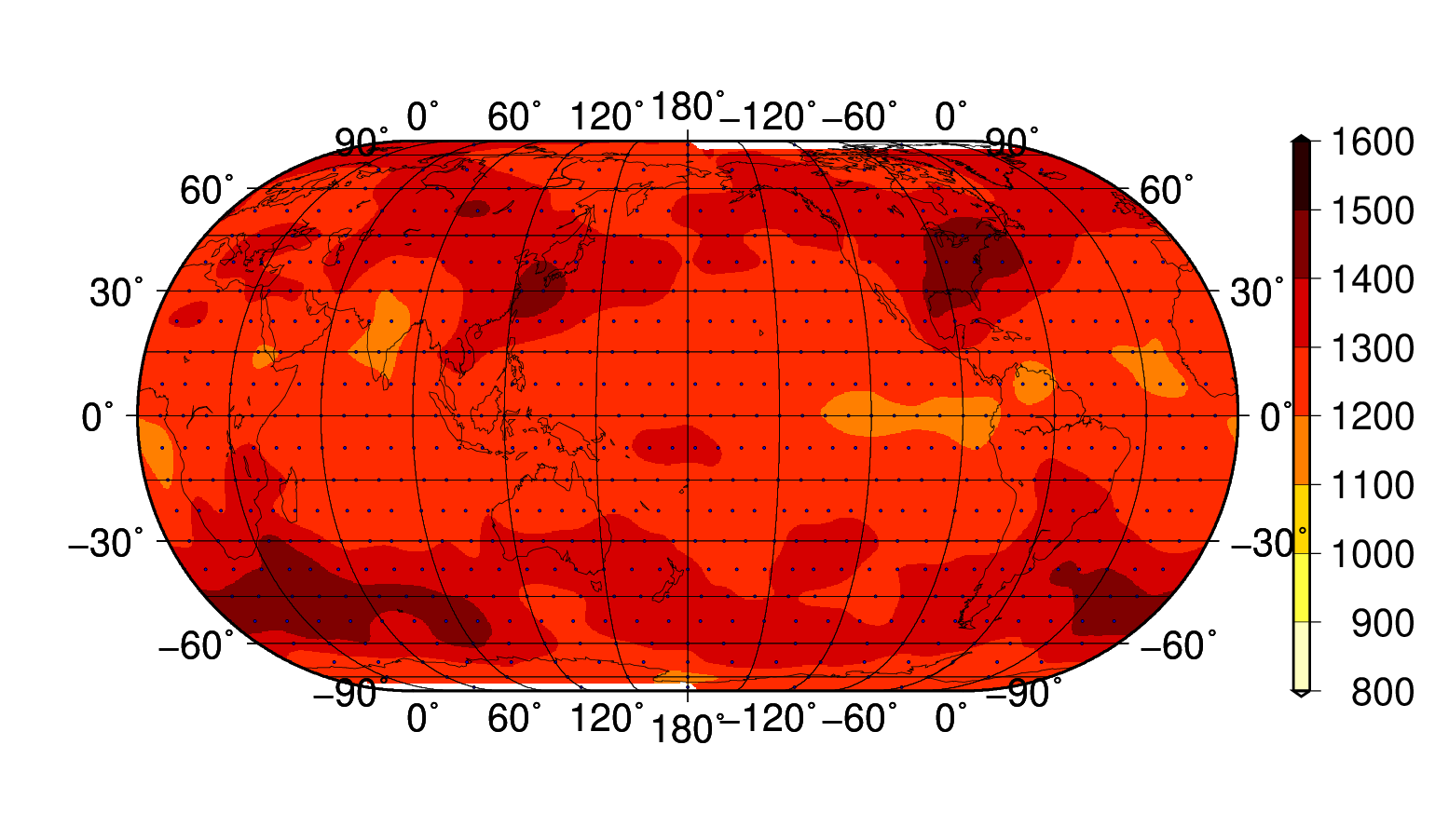}}
\caption{\label{fig:inoutdistnonen}Spatial distribution of the weighted degrees of nodes (\textit{in} and \textit{out}) averaged over non-El-Ni\~{n}o times. The nodes of the network are shown by small dots.}
\end{figure}

Next we show in Figs.~\ref{fig:inoutdistnonen}\subref*{fig:indistnonen},\subref*{fig:outdistnonen} the spatial distribution of $I^{y}_{l}$ and $O^{y}_{l}$ during normal (non-El-Ni\~{n}o) times (We used the GMT tool ~\cite{gmt} to overlay the fields on maps). Our statement about the \textit{in--}weighted degrees having a broader distribution by 10 to 20 percent at all times, gets its direct spatial interpretation. As seen in Fig.~\ref{fig:inoutdistnonen}, the \textit{out--} weighted degree spatial field is much more homogeneous compared to the \textit{in--} weighted degree which has a sharp minimum around ENB. This asymmetry between $I_{\bf C}^{y}$ and $O_{\bf C}^{y}$ becomes significantly more pronounced during El-Ni\~{n}o periods (see Fig.~\ref{fig:inoutdisten}). The cluster ${\bf C}$ is marked explicitly with large blue circles. In this sense, ENB gets significantly more autonomous during El-Ni\~{n}o. A full time dependent animation of these maps is available at ~\cite{prl2_pg}. Note that the fields in Figs.~\ref{fig:inoutdistnonen} and ~\ref{fig:inoutdisten} have been calculated without any correlation threshold.


In summary, we have found a new dynamical pattern that reflects the coupling between the El-Ni\~{n}o basin, and the rest of the world. ENB becomes significantly more autonomous during El-Ni\~{n}o, losing a large fraction of its \textit{in}--links, while still having \textit{out}--links. This kind of topology is reminiscent of pacemakers in network models~\cite{scc}. The major impact of events inside ENB on world climate on one hand, and the weakened correlations during El-Ni\~{n}o episodes on the other hand~\cite{survival,tsoniselnino,EPLgozyamga}, are thus not contradicting. In fact, the uni--directional interaction of ENB with large parts of the climate network might suggest the origin for its significant dynamical role in the global climate. 

Our results also suggest the existence of a robust delayed relationship between the inward and outward coupling of ENB with the rest of the world. The emergence of the autonomous behavior is consistently following, after typically 3 months, a short and weak episode of a decreased outward coupling. The two fields (outward and inward coupling) oscillate with a phase shift. 

We also find that ENB is forced more by the northern hemisphere than by the southern, and forces the southern hemisphere more than it forces the northern. Since the annual cycle in the two hemispheres is opposite, this north--south asymmetry might be related to the known (not yet fully understood) partial phase locking of the ENSO cycle with the annual cycle (see e.g.~\cite{mghilscience,tdelaytziperm}). 

\textbf{Acknowledgements} We thank the EU EPIWORK project, the Israel Science Foundation, DTRA, ONR, and Research project for a sustainable development of economic and social structure dependent on the environment of the eastern coast of Asia.


\begin{thebibliography}{21}
\bibitem[{fn1()}]{fn1}
\bibinfo{note}{The standard deviation of a weighted degree over time is
  typicaly between 13 and 16 percent of its mean value.}
\bibitem[{fn2()}]{fn2}
\bibinfo{note}{We chose $T$ as the highest possible value, such that zones that
  do not belong to ENB are not part of ${\bf C}$.}
\bibitem[{fn3()}]{fn3}
\bibinfo{note}{During major events we can even see a full second oscillation in
  which the $I$ field recovers, and $O$ weakens again, giving rise to a second
  major weakening of $I$.}
\bibitem[{fn5()}]{fn5}
\bibinfo{note}{The origin of error is due to low sampling rates since we take a
  snapshot of the network every 50 days}.
\end{thebibliography}
\end{document}